# Possible Competition between superconductivity and magnetism in $RuSr_2Gd_{1.5}Ce_{0.5}Cu_2O_{10-\delta}$ (Ru-1222) Rutheno-cuprate compounds

V.P.S. Awana[*], M.A. Ansari, Anurag Gupta, R.B. Saxena, and H. Kishan

National Physical Laboratory, K.S. Krishnan Marg,

New Delhi 110012, India

and

Devendra Buddhikot and S.K. Malik

Tata Institute of Fundamental Research, Homi Bhabha Road,

Mumbai 400005, India

The $RuSr_2Gd_{1.5}Ce_{0.5}Cu_2O_{10-\delta}$ (Ru-1222) compounds, with varying oxygen content, crystallize in a tetragonal crystal structure (space group $I4/mmm$). Resistance (R) versus temperature (T) measurements show that the air-annealed samples exhibit superconductivity with superconducting transition temperature ($T_c$) onset at around 32 K and R=0 at 3.5 K. On the other hand, the $N_2$-annealed sample is semiconducting down to 2 K. Magneto-transport measurements on air-annealed sample in applied magnetic fields of 3 and 6 Tesla show a decrease in both $T_c$ onset and $T_{R=0}$. Magnetoresistance of up to 20% is observed in $N_2$-annealed sample at 2 K and 3 T applied field. The DC magnetization data (M vs. T) reveal magnetic transitions ($T_{mag.}$) at 100 K and 106 K, respectively, for both air- and $N_2$-annealed samples. Ferromagnetic components in the magnetization are observed for both samples at 5K and 20 K. The superconducting transition temperature ($T_c$) seems to compete with the magnetic transition temperature ($T_{mag.}$). Our results suggest that the magnetic ordering temperature ($T_{mag.}$) of Ru moments in $RuO_6$ octahedra may have direct influence/connection with the appearance of superconductivity in $Cu-O_2$ planes of Ru-1222 compounds.

# 1. INTRODUCTION

Recent discovery of the coexistence of magnetism and superconductivity in Ru-1222 ($RuSr_2(Gd,Ce)_2Cu_2O_{10-d}$) [1,2] and Ru-1212 ($RuSr_2GdCu_2O_{8-d}$) [3-6] Rutheno-cuprates compounds has attracted a great deal of attention. Both Ru-1212 and Ru-1222 compounds are related with the Cu-1212 ($CuBa_2RECu_2O_7$ or RE-123, RE = rare earths) compounds with Ba in the latter replaced by Sr and oxygen deficient $CuO_{1-d}$ chains replaced by $RuO_6$ octahedra. Further, the oxygen free RE layer between $CuO_2$ planes in Cu-1212 is replaced by a rocksalt O-(Gd,Ce)-O block in Ru-1222 structure. Synthesis of both Ru-1212 and Ru-1222 compounds in pure phase has been a problem due to the formation of magnetic $SrRuO_3$ and $Gd_2SrRuO_6$ in the matrix [1-6].

By now, it is widely accepted [1,2, 4-7] that the Ru moments in $RuO_6$ octahedra of Ru-1212/Ru-1222 order magnetically around 100-140K ($T_{mag.}$), though the exact nature of ordering is still debated [8-10]. Further, superconductivity associated with the $CuO_2$ planes with transition temperature ($T_c$) of up to 40 K [1-10] is observed in these compounds. Co-existence of superconductivity and magnetism in these compounds has been indirectly proved by nuclear magnetic resonance (NMR) experiments [11]. At the same time, it is worth mentioning that there exist some reports, which are against the genuine co-existence of superconductivity with magnetism in these compounds [12,13].

Both, the magnetic ordering temperature ($T_{mag.}$) of the Ru moments and the superconductivity ($T_c$) of the $CuO_2$ planes are expected to be controlled by the extent of $Ru^{4+}/Ru^{5+}$ ratio in $RuO_6$ octahedra. The controlling of $T_{mag}$ is due to mixed $Ru^{4+}/Ru^{5+}$ ratio and that of the $T_c$ is due to the doping of $CuO_2$ superconducting planes by charge transfer from $RuO_6$ octahedra. Oxygen content of $RuO_6$ octahedra or the ratio of $Ru^{4+}/Ru^{5+}$ can be controlled to some extent in Ru-1222, but not in Ru-1212 [14]. Therefore, it is of prime interest to vary the oxygen content of $RuO_6$ octahedra, and study its effect on the $T_{mag.}$ and $T_c$ of the Ru-1222 compound. This is the aim of our present work. We have been successfully in controlling the $T_c$ of Ru-1222 compound from 3.5K to non-superconducting within the same crystallographic phase. This provided us an opportunity to compare the $T_c$ and $T_{mag}$ of Ru-1222. Our results indicated that both $T_c$ and $T_{mag}$ compete with each other.

## 2. EXPERIMENTAL DETAILS

The $RuSr_2Gd_{1.5}Ce_{0.5}Cu_2O_{10-\delta}$ (Ru-1222) samples were synthesized through a solid-state reaction route from $RuO_2$, $SrCO_3$, $Gd_2O_3$, $CeO_2$ and CuO. Calcinations were carried out on the mixed powder at 1000, 1020, 1040 and 1050ºC each for 24 hours with intermediate grindings. The pressed circular pellets were annealed in air for over 48 hours at 1050ºC and are named "air-annealed". One of the air-annealed pellets was further annealed in nitrogen gas (1 atm) at 420°C for 24 hours and subsequently cooled slowly to room temperature (named "$N_2$-annealed"). X-ray diffraction (XRD) patterns were obtained at room temperature (MAC Science: MXP18VAHF[22]; Cu$K_\alpha$ radiation). Magnetization measurements were performed on a SQUID magnetometer (Quantum Design: MPMS-5S). Resistivity measurements under applied magnetic fields of up to 6 T were made in the temperature range of 2 to 300 K using a four-point-probe technique (PPMS Quantum Design).

## 3. RESULTS AND DISCUSSION
### a): X-ray diffraction and Phase formation

Both air- and $N_2$-annealed Ru-1222 compounds crystallize in a tetragonal structure (space group *I4/mmm*) with $a = b = 3.8427(7)$Å and $c = 28.4126(8)$ Å for the former and $a = b = 3.8498(3)$ Å and $c = 28.4926(9)$ Å for the latter. The increase in the lattice parameters of $N_2$-annealed sample indicates an overall decrease in oxygen content of the sample. X-ray diffraction patterns for both the air- and $N_2$-annealed samples are shown in Fig. 1. Small impurity peaks (marked with "*") are seen close to the background. Presence of small amounts of $SrRuO_3$ and/or $GdSr_2RuO_6$ in Ru-1222 samples has been noted earlier also [1,2,5]. Our currently studied samples are, in fact, far better in terms of their phase purity, compared to those reported earlier by various authors. As mentioned earlier, Ru-1222 is structurally related to the Cu-1212, e.g. $CuBa_2YCu_2O_{7-\delta}$ phase with Cu in the charge reservoir in the latter replaced by Ru such that the Cu-O chain is replaced by a $RuO_2$ sheet. Furthermore, a three-layer fluorite-type block, instead of a single oxygen-free $R$ (R= rare earth element) layer, is inserted between the two $CuO_2$ planes of the Cu-1212 structure [15] to get Ru-1222 phase.

**b): Electrical transport under magnetic field and Magnetism**

The resistance versus temperature (R-T) behavior of air-annealed Ru-1222 sample in magnetic fields of 0, 3 and 6 Tesla in the temperature range of 2-300 K is shown in Fig. 2. The resistance of the compound increases with decrease in temperature indicating its semiconducting behavior. However, in zero applied field, the compound shows a sharp drop in its resistance at around 35 K ($T_c^{onset}$) with zero resistance state occurring at 3.5 K ($T_c^{(R=0)}$). Rather broad superconducting transition and the low value of $T_c$ (3.5 K) point towards low carrier concentration in the air-annealed sample. It is well known that superconducting properties of Ru-1222 can be considerably improved by annealing in oxygen under pressure [10,16]. In fact, by annealing in 100-atm.-$O_2$, one can achieve $T_c$ of up to 40-45 K [10]. By normal pressure (1 atm.) $O_2$ annealing, the $T_c$ of the air-annealed sample could not be increased significantly.

In applied magnetic fields of 3 and 6 Tesla, though the normal state (above $T_c$ onset) resistance behavior of Ru-1222 is essentially the same as in zero field, both $T_c^{onset}$ and $T_c^{(R=0)}$ decrease with applied field. For example, $T_c^{onset}$ is 32 K, 27 K and 15 K, respectively for 0, 3 and 6 Tesla applied fields and $T_c^{(R=0)}$ state is not observed under these applied magnetic fields. Magneto-resistance, (MR), defined as,

$$MR\% = [(R_H - R_0)/R_H] \times 100 \qquad (1).$$

is plotted in inset of Fig.2. Though the observed MR% is small, it changes sign at around 180 K. being positive above 180K and negative below 180K. The sign change of MR at 180 K might be related to the magnetic structure transformation at the said temperature. We shall discuss this later after presenting magnetization results.

Figure 3 depicts the resistance versus temperature (R-T) behavior for $N_2$-annealed Ru-1222 sample in magnetic fields of 0, 3 and 6 Tesla. The R-T behavior of this compound is semiconducting down to 2 K. No superconducting transition is observed in the whole temperature range studied (2-300 K). Further, in low temperature region, an appreciable MR is seen in this sample. Magnetoresistance, as a function of applied field, at temperatures of 5 and 10 K, is plotted in inset of Fig. 3. MR of >20 % is observed at 5 K in an applied field of up to 9 Tesla. At 2 K, around 20% MR is seen even in low applied field of 3 Tesla (plot not shown).

Figure 4 shows the magnetic susceptibility ($\chi$) vs. temperature (T) behaviour in the temperature range of 2 to 300 K for air-annealed Ru-1222 sample in an applied field of 100 Oe, measured in both zero-field-cooled (ZFC) and field-cooled (FC) modes. The $\chi$ vs. T plot shows the branching of ZFC and FC curves at around 95 K. and the magnetic susceptibility starts shooting up at around 100 K ($T_{mag}$). In fact, the susceptibility of Ru-1222 compound starts deviating from normal Curie-Weiss paramagnetic behaviour at around 165 to 200 K. It has been reported that the Ru moments in Ru-1222 order antiferromagnetically at around 180 K, which later develops in to a canted ferromagnetism at lower temperatures ($T_{mag.}$) [1,2,6]. As mentioned earlier, the exact nature of magnetic ordering in low temperature region is still debated. The 180 K paramagnetic to antiferromagnetic transition could be seen in magneto-transport measurements also, where small positive MR changes to negative MR below this temperature (inset, Fig. 2).

The characteristic temperature $T_{mag}$ is weakly dependent on applied magnetic field H < 100 Oe. For H > 1000 Oe, both ZFC and FC magnetization curves are merged with each other (see inset, Fig. 4). In fact, no ZFC - FC branching is observed down to 2 K in 5000 Oe field. This is in general agreement with earlier reports [1,2]. Superconductivity is not seen in terms of diamagnetic transition ($T_d$) in applied field of H = 100 Oe (Fig.4). It is known that, due to internal magnetic fields, these compounds are in a spontaneous vortex phase (SVP) even in zero external field [17]. For $T_d < T < T_c$, the compound remains in a mixed state. Hence though R= 0 is achieved at relatively higher temperatures (3.5 K, see Fig. 2), the diamagnetic response is seen at much lower temperatures and that too in quite small applied magnetic ($H_{c1}$ < 25 Oe) fields. Hence we could conclude that the magnetic susceptibility ($\chi$) vs. T behaviour shown in Fig. 4 does not exclude the presence of superconductivity in the presently studied air-annealed Ru-1222 sample. At lower applied field of 5 Oe, the compound exhibits diamagnetic transition ($T_d$) below $T_c^{(R=0)}$ state (curve not shown).

The $\chi$-T behaviour in the temperature range of 2 to 300 K for $N_2$-annealed Ru-1222 sample in an applied fields of 100 Oe, measured in both zero-field-cooled (ZFC) and field-cooled (FC) modes, is shown in Fig. 5. The general shape of FC and ZFC magnetization plots is similar to that for Air-annealed sample. The interesting change is that $T_{mag.}$ (defined earlier) has increased to 106 K for $N_2$-annealed sample. It is worth mentioning that the $N_2$-annealed sample is not superconducting down to 2 K (Fig. 3, R vs. T results).

### c): Ferromagnetic component

Magnetization (M) versus applied field (H) isotherm at 5 K for air-annealed Ru-1222 sample is shown in Fig. 6. The magnetization starts saturating above 6 Tesla field in both directions. The M-H plot is further zoomed in applied field of –900 Oe to 900 Oe, and shown in inset of Fig. 6. At 5 K, the returning moment ($M_{rem}$) i.e. the value of magnetization at zero returning field and the coercive filed ($H_c$), i.e. the value of applied returning field needed to get zero magnetization, are respectively 2.50 emu/gram and 160 Oe (inset Fig. 6). It is known that the Gd (magnetic rare earth) in this compound orders magnetically below 2 K and Ce remains in a tetravalent, non-magnetic state, hence the $M_{rem}$ and $H_c$ arising from the ferromagnetic hysteresis loops do belong to Ru only. Interestingly, for Ru-1212, the hysteresis loops are reported to be quite narrow [4,5]. This indicates that, in Ru-1222, the ferromagnetic domains are less anisotropic and more rigid. The M-H loop at 20 K for air-annealed Ru-1222 sample is shown in Fig. 7, which shows a decreased $M_{rem}$ (1emu/gram) and $H_c$ (75 Oe) compared to 5 K plot in Fig. 6. The values of both $M_{rem}$ and $H_c$ decrease with T. Hysteresis loops were not seen at higher temperatures above $T_{mag}$. For example, the M-H plot at 150 K is seen as completely linear with field (Fig. 8), and no hysteresis loops were visible even after zooming at low fields (inset of Fig. 8).

Figure 9 depicts the M-H plot at 5 K for $N_2$-annealed Ru-1222 sample. This plot is similar to that observed for the air-annealed Ru-1222 sample. The zoomed ferromagnetic component is shown in the inset. The interesting difference, when compared with air-annealed sample, is that M does not saturate in applied fields up to 7 Tesla. This is in contrast to the M-H plot for air-annealed sample at 5 K for which M saturates in a field of 6 Tesla (see Fig. 6). The M-H plot for $N_2$-annealed sample is further zoomed in applied field of –900 Oe to 900 Oe, and shown in the inset of Fig. 9. Ferromagnetic loop is seen clearly with $M_{rem}$ (2emu/gram) and $H_c$ (170Oe). The M-H loop at 20 K for $N_2$-annealed sample is shown in Fig. 10, which reveals decreased $M_{rem}$ (1emu/gram) and $H_c$ (70 Oe). The trend of $M_{rem}$ and $H_c$ for $N_2$-annealed sample is similiar to that of Air-annealed sample, i.e both decrease with increase in temperature. The M-H loops were not seen above $T_{mag}$ for the $N_2$-annealed samples also.

The results presented above for both air- and $N_2$-annealed presently studied Ru-1222 samples may be summarised as follows:

1. Both air- and $N_2$-annealed samples crystallize in a single-phase tetragonal structure (space group *I4/mmm*).

2. The air-annealed sample is superconducting with $T_c^{(R=0)}$ of 3.5 K, while the $N_2$-annealed is not superconducting down to 2 K. The $N_2$-annealed sample exhibits negative magneto-resistance of up to 20% at 5 K.

3. Both air- and $N_2$-annealed samples order magnetically, with $T_{mag}$ of nearly 100 K for the former and 106 K for the latter.

4. Both air- and $N_2$-annealed samples exhibit ferromagnetic loops below $T_{mag}$. The relative width of loops is more (higher $M_{rem}$ and $H_c$) for air-annealed sample compared to that of $N_2$-annealed sample at a fixed temperature.

5. For air-annealed sample, the M-H behaviour at 5 K exhibits near saturation above 6 Tesla. However, for $N_2$-annealed sample, M is not saturated up to applied field of 7 Tesla.

We now discuss these results point by point. As far as the phase formation and crystallization of the two Ru-1222 samples is concerned, it seems that there is some scope for oxygen to get released from $RuO_6$ octahedra while still maintaining its crystal structure. In fact, another school of thought also exists, which believes in the release of oxygen from rocksalt O-(Gd,Ce)-O block and not from $RuO_6$ in Ru-1222. Recent spectroscopic studies on Ru-1222 do show indications towards valence fluctuations of $Ru^{4+}/Ru^{5+}$ and hence some variation in oxygen content of $RuO_6$ octahedra [18]. It has been reported earlier that oxygen content of Ru-1222 is tuneable to some extent by $N_2/Ar_2$ annealing but not for Ru-1212 [14]. Thus it has been possible for us, in the present study, to get both air- and $N_2$-annealed Ru-1222 samples in the same crystal structure. Decreased oxygen content of $N_2$-annealed sample is indicated by increased c-parameter of the sample (the exact oxygen content has not been determined).

As far as the second point is concerned, the air-annealed sample is superconducting but the $N_2$-annealed sample is not. Further, the normal state conduction, though semiconducting for both samples, is relatively better for air-annealed sample. This indicates that the air-annealed sample has relatively more mobile hole-carriers than the $N_2$-annealed sample. The doping of mobile holes in widely believed conducting/superconducting Cu-$O_2$ planes in RE-123 compounds takes place by charge transfer from oxygen variable redox $CuO_{1-d}$ chains. In Ru-1222, the role of redox layer is supposedly played by $RuO_6$ octahedra. In $N_2$-annealed sample, the $Ru^{4+}/Ru^{5+}$ ratio will be different than that in air-annealed sample, and hence the number of transferred mobile holes to Cu-$O_2$ planes will be different. As far as magneto-transport is concerned, no appreciable MR is seen above $T_{mag}$ for both the samples. For air-annealed superconducting sample, the broadening of transition under field is seen and is similar to that in other HTSC compounds. For $N_2$-annealed sample, negative MR of around 20% is seen at 2 K. As we know from present magnetization measurements and various other earlier reports, that the low temperature magnetism of Ru-1222 is complex and calls for magnetic phase separation including spin glass [10], canted antiferromagnetism [19], or both simultaneously. Though magnetic structure of Ru-1222 in not yet revealed by low temperature neutron scattering measurements, one thing is apparent that the compound possesses complex low temperature magnetism. In such situations, appreciable MR is expected due to tunnelling between various magnetic domains arising from magnetically phase separated system. The $N_2$-annealed sample is highly under-doped and hence one presumes that electrical transport conduction is mainly through $RuO_6$ layer, where magnetic scattering does take place. The role of magnetic charge spin scattering has earlier been observed in Ru-1212, where $T_{mag}$ is seen clearly as a hump in resistivity measurements [4,12]. To our knowledge this is the first observation of finding appreciable negative MR at 2K in a non-superconducting Ru-1222 compound. This is in conformity with the earlier reports of magnetic phase separation at low temperature in Ru-1222 [10,19].

Third point comprises of the fact that, for superconducting air-annealed sample, $T_{mag}$ is 100 K while the same is 106 K for non-superconducting $N_2$-annealed sample. As the $T_c$ (superconducting transition temperature) increases, the $T_{mag.}$ (magnetic transition temperature of Ru) decreases. This is further demonstrated by the fact that 100-atm-$O_2$ post annealed sample possesses a $T_c$ of 43 K along with $T_{mag.}$ of around 90 K [21]. $T_{mag}$ depends weakly on the measuring DC magnetic field. Hence one should be careful in comparing the $T_{mag.}$ and $T_c$ of various Ru-1222 samples. Also one should be reminded of the fact that the decreased number of mobile carriers in Cu-$O_2$ planes of $N_2$-annealed sample along with other structural parameters may also be partly responsible for the non superconducting behavior. One cannot explicitly say that increased $T_{mag.}$ is responsible for the decreased superconductivity of the system. What is apparent though, is the possible competing nature of $T_{mag.}$ and $T_c$.

As discussed in point 1 above, the $Ru^{4+}/Ru^{5+}$ ratio in variously processed Ru-1222 samples is supposed to be different. The amount of $Ru^{4+}$ will be higher in $N_2$-annealed sample [14,18]. $T_{mag}$ originates from the ordering of Ru moments in $RuO_6$ octahedra. Changed amount of $Ru^{4+}/Ru^{5+}$ magnetic spins contribution is responsible for different $T_{mag}$ of the two compounds. Higher content of $Ru^{4+}$ gives rise to increased $T_{mag}$.

Fourth point says that though both compounds have ferromagnetic domains at low temperatures below $T_{mag}$, the characeresitc values of $M_{rem}$ and $H_c$ are relatively higher for air-annealed sample. Our detailed micro-structural studies earlier for Ru-1222 showed that the observed super-lattice structures due to tilt of $RuO_6$ octahedra [20] might be coupled with the weak ferromagnetic domains constructed by ordering of the canted Ru moments below the magnetic transition temperature ($T_{mag}$). Hence coupling of ferromagnetic domains depends on the long range ordering of tilted $RuO_6$ octahedras in a given Ru-1222 system. In $N_2$-annealed sample, the long-range superstructures may break down relatively at smaller length scale than in air-annealed sample due to less oxygen in $RuO_6$ octahedra of the same giving rise to weak coupling of the ferromagnetic domains. This will give rise to lower values of $M_{rem}$ and $H_c$. This also explains the fifth point regarding the observed saturation of M-H curve for air-annealed sample and not for $N_2$-annealed sample. The saturation of M-H is dependent on the long range coupling of aligned ferromagnetic domains, which is observed for air-annealed sample only. Long range coupling of aligned moments is directly dependent on the stability of $RuO_6$ octahedra tilt angle superstructures, which is certainly less for $N_2$- annealed sample due to break down in homogenous oxygen content close to 6.0 in the octahedra. In earlier reports also, the M-H plots at 5 K for superconducting Ru-1222 samples, saturated below an applied field of 7 Tesla [1,2,16,21].

## 6. SUMMARY AND CONCLUSIONS

The Ru-1222 Rutheno-cuprate compounds have been synthesized with both magneto-superconducting (air-annealed) and only magnetic ($N_2$-annealed) properties. Same crystal structure is maintained for both compounds. The Ru magnetic ordering temperature ($T_{mag.}$) is higher for non-superconducting sample than for the superconducting sample. M-H plots at 5 K saturate below 6 Tesla applied field for superconducting sample, but not for non-superconducting sample. These results could be explained on the basis of the presence of mixed valency of $Ru^{4+}/Ru^{5+}$ in these compounds. It is apparent that magnetic order of Ru in $RuO_6$ octahedra and superconductivity ($T_c$) in $Cu-O_2$ planes are intimately related with each other, indicating the co-existance of magntism and superconductivity in Ru-1222 compounds.

**FIGURE CAPTIONS**

Figure 1. Observed X-ray diffraction pattern for the air- and $N_2$ – annealed $RuSr_2Gd_{1.5}Ce_{0.5}Cu_2O_{10-\delta}$ samples.

Figure 2. Resistance (R) vs. temperature (T) plots in 0, 3, and 6 T applied magnetic fields for air-annealed $RuSr_2Gd_{1.5}Ce_{0.5}Cu_2O_{10-\delta}$ sample. Inset shows the Magnetoresistance (MR%) vs. T plot for 6 Tesla applied field.

Figure 3. Resistance (R) vs. temperature (T) plots in 0, 3, and 6 T applied magnetic fields for $N_2$-annealed $RuSr_2Gd_{1.5}Ce_{0.5}Cu_2O_{10-\delta}$ sample. Inset shows the Magnetoresistance (MR%) at 5 K in applied fields up to 9 Tesla.

Figure 4. Magnetic susceptibility ($\chi$) vs. temperature (T) plot for air–annealed $RuSr_2Gd_{1.5}Ce_{0.5}Cu_2O_{10-\delta}$ in both ZFC and FC modes with applied field of 100 Oe, inset shows the same for H = 5000 Oe.

Figure 5. Magnetic susceptibility ($\chi$) vs. temperature (T) plot for $N_2$ – annealed $RuSr_2Gd_{1.5}Ce_{0.5}Cu_2O_{10-\delta}$ in both ZFC and FC modes with applied field of 100 Oe.

Figure 6. Magnetization-Field (M–H) hysteresis loop for air-annealed $RuSr_2Gd_{1.5}Ce_{0.5}Cu_2O_{10-\delta}$ at 5 K. Inset shows the same for –900Oe = H = +900Oe.

Figure 7. Magnetization-Field (M–H) hysteresis loop for air-annealed $RuSr_2Gd_{1.5}Ce_{0.5}Cu_2O_{10-\delta}$ at 20 K Inset shows the same for –900Oe = H = +900Oe.

Figure 8. Magnetization-Field (M–H) hysteresis loop for air-annealed $RuSr_2Gd_{1.5}Ce_{0.5}Cu_2O_{10-\delta}$ sample at 150 Inset shows the same for –900Oe = H = +900Oe.

Figure 9. Magnetization-Field (M–H) hysteresis loop for $N_2$-annealed $RuSr_2Gd_{1.5}Ce_{0.5}Cu_2O_{10-\delta}$ sample at 5 K Inset shows the same for –900Oe = H = +900Oe.

Figure 10. Magnetization-Field (M–H) hysteresis loop for $N_2$-annealed $RuSr_2Gd_{1.5}Ce_{0.5}Cu_2O_{10-\delta}$ sample at 20 K. Inset shows the same for –900Oe = H = +900Oe.

**REFERENCES**

1. I. Felner, U. Asaf, Y. Levi, and O. Millo, Phys. Rev. B 55, R3374 (1997)

2. I. Felner, and U. Asaf, Int. J. Mod. Phys. B. 12, 3220 (1998)

3. L. Bauernfeind, W. Widder and H.F. Braun, Physica C 254, 151 (1995).

4. C. Bernhard, J.L. Tallon, Ch. Niedermayer, Th. Blasius, A. Golnik, E. Brücher, R.K. Kremer, D.R. Noakes, C.E. Stronack, and E.J. Asnaldo, Phys. Rev. B 59, 14099 (1999).

5. J. L. Tallon, J. W. Loram, G.V.M. Williams, and C. Bernhard, Phys. Rev. B 61, R6471 (2000)

6. V.P.S. Awana, S. Ichihara, J. Nakamura, M. Karppinen, and H. Yamauchi, Physica C 378-381, 249-254 (2002).

7. A. Butera, A. Fainstein, E. Winkler, and J. Tallon, Phys. Rev. B 63, 54442 (2001).

8. J.W. Lynn, B. Keimer, C. Ulrich, C. Bernhard, and J.L. Tallon, Phys. Rev. B 61, R14964 (2000).

9. V.P.S. Awana, T. Kawashima and E. Takayama-Muromachi, Phys. Rev. B. 67, 172502 (2003).

10. C. A. Cardoso, F.M. Araujo-Moreira, V.P.S. Awana, E. Takayama-Muromachi, O.F. de Lima, H. Yamauchi, M. Karppinen, Phys. Rev. B 67, 024407R (2003).

11. Y. Tokunaga, H. Kotegawa, K. Ishida, Y. Kitaoka, H. Takigawa, and J. Akimitsu, Phys. Rev. Lett. 86, 5767 (2001).

12. V.P.S. Awana, S. Ichihara, J. Nakamura, M. Karppinen, H. Yamauchi, Jinbo Yang, W.B. Yelon, W.J. James and S.K. Malik, J. Appl. Phys. 91, 8501 (2002).

13. C.W. Chu, Y.Y. Xue, S. Tsui, J. Cmaidalka, A.K. Heilman, B. Lorenz, and R.L. Meng, Physica C 335, 231 (2000).

14. M. Matvejeff, V.P.S. Awana, H. Yamauchi and M. Karppinen, Physica C 392-396, 87-92 (2003).

15. N. Sakai, T. Maeda, H. Yamauchi and S. Tanaka, Physica C 212, 75 (1993).

16. M.T. Escote, V.A. Meza, R.F. Jardim, L. Ben-Dor, M.S. Torikachvill and A.H. Lacerda, Phys. Rev. B. 66, 144503 (2002).

17. E.B. Sonin, and I. Felner, Phys. Rev. B 57, R14000 (1998).

18. V.P.S. Awana, M. Karppinen, H. Yamauchi, R.S. Liu, J.M. Chen and L.-Y. Jang, J. Low Temp. Physics 131, 1211 (2003).

19. I. Živkovic, Y. Hirai, B.H. Frazer, M. Prester, D. Drobac, D. Ariosa, H. Berger, D. Pavuna, G. Margaritondo, I. Felner and M. Onellion, Phys. Rev. B 65, 144420 (2002).

20. Tadahiro Yokosawa, Veer Pal Singh Awana, Koji Kimoto, Eiji Takayama-Muromachi, Maarit Karppinen, Hisao Yamauchi and Yoshio Matsui, Ultramicroscopy 98, 283-295 (2004).

21. V.P.S. Awana, E. Takayama-Muromachi, M. Karppinen, and H.Yamauchi, Physica C 390, 233-238 (2003).

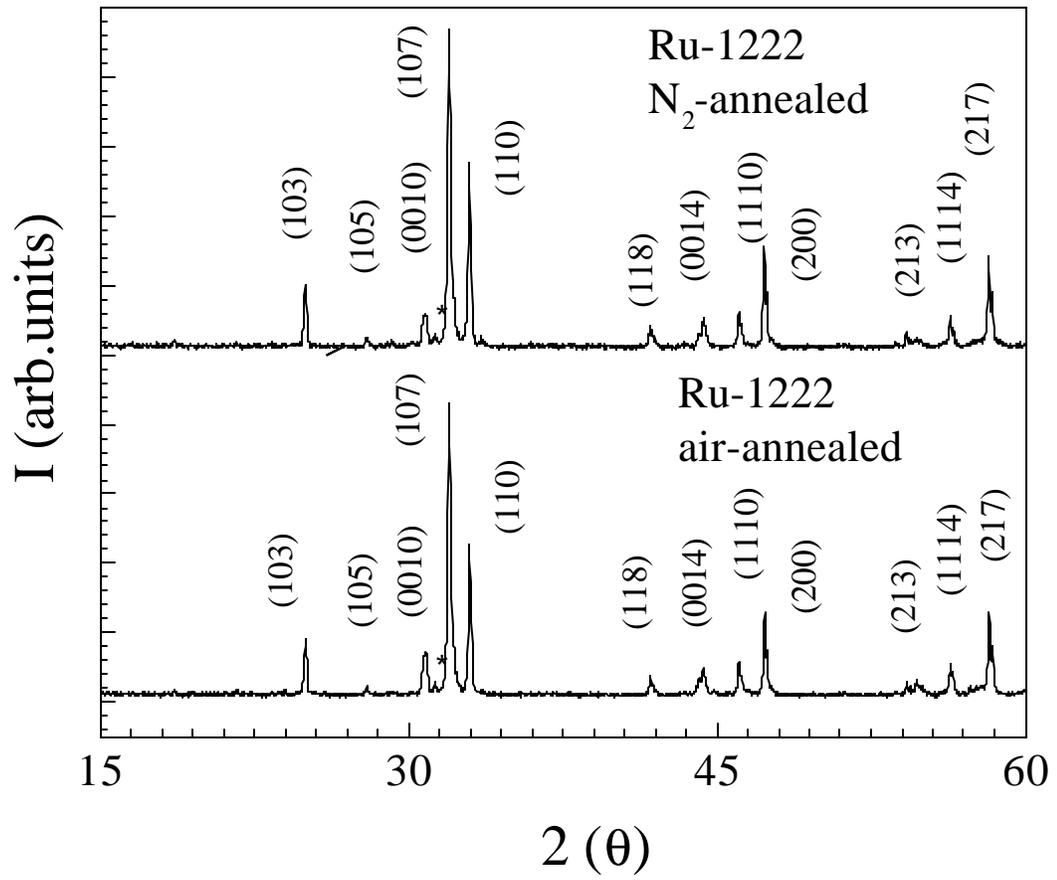

Fig.1 Awana etal.

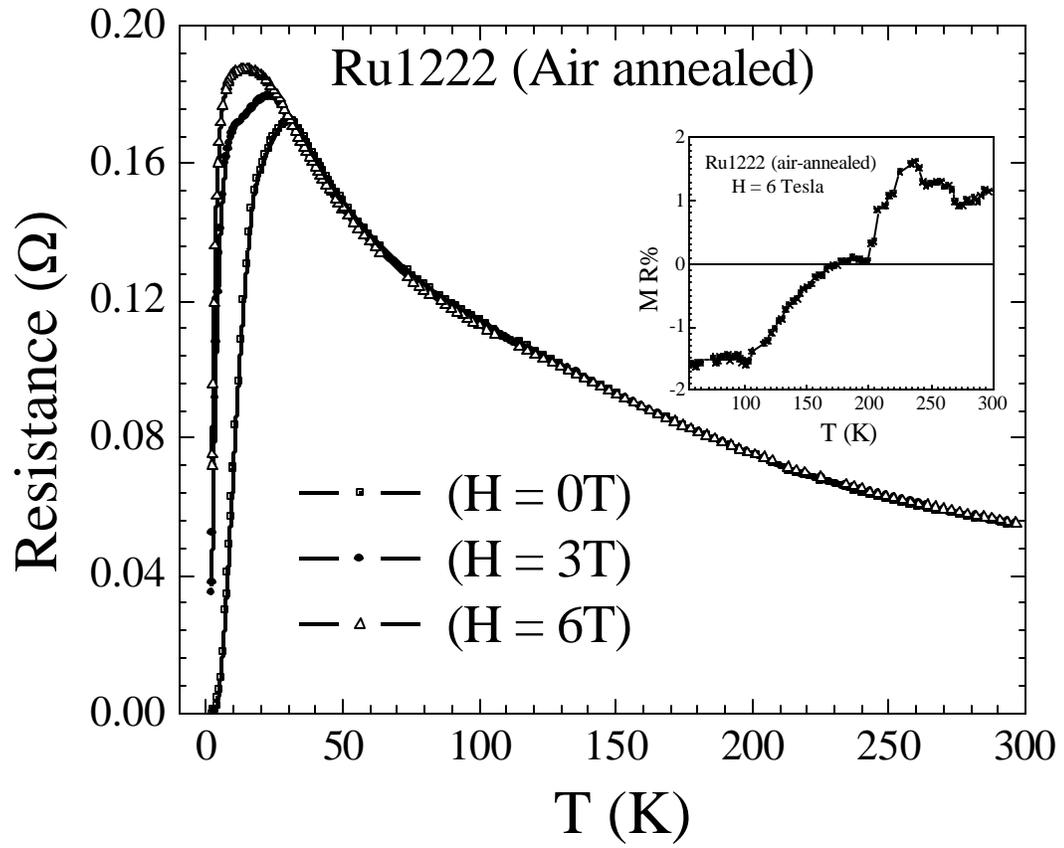

Fig.3 Awana et al.

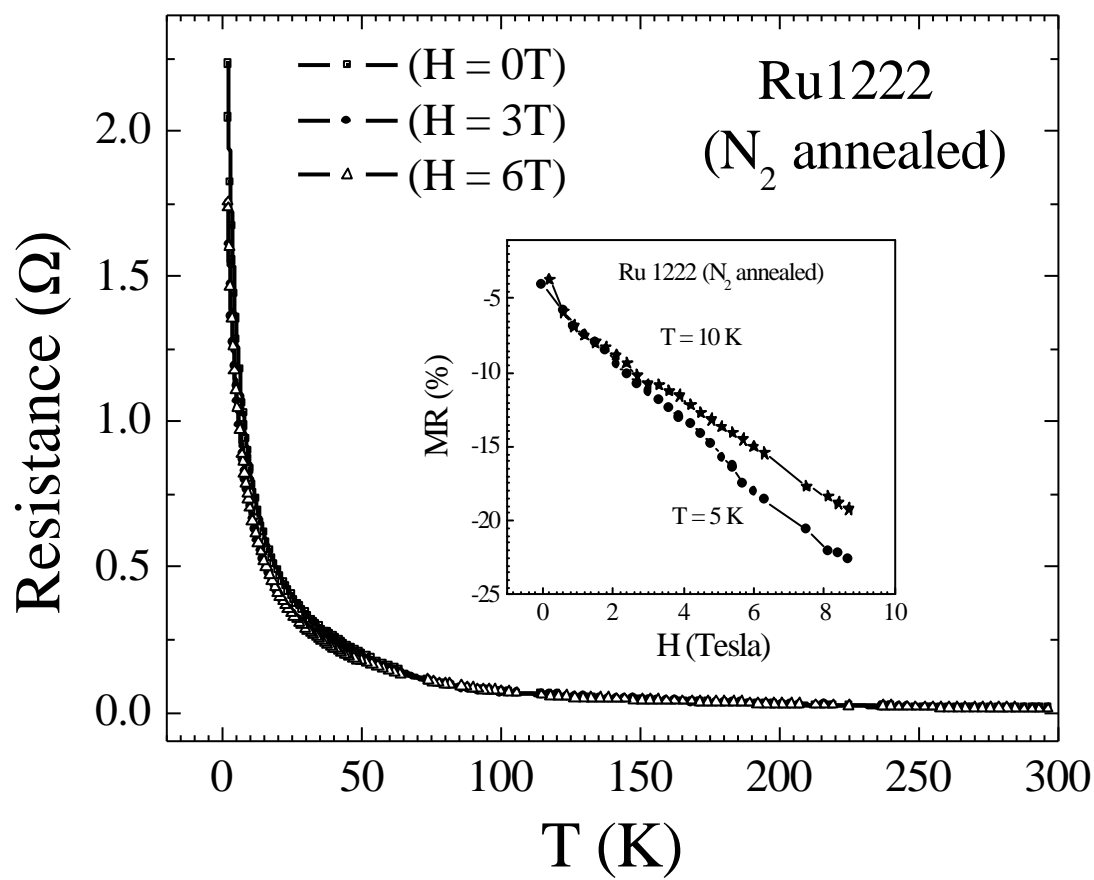

Fig.4 Awana etal.

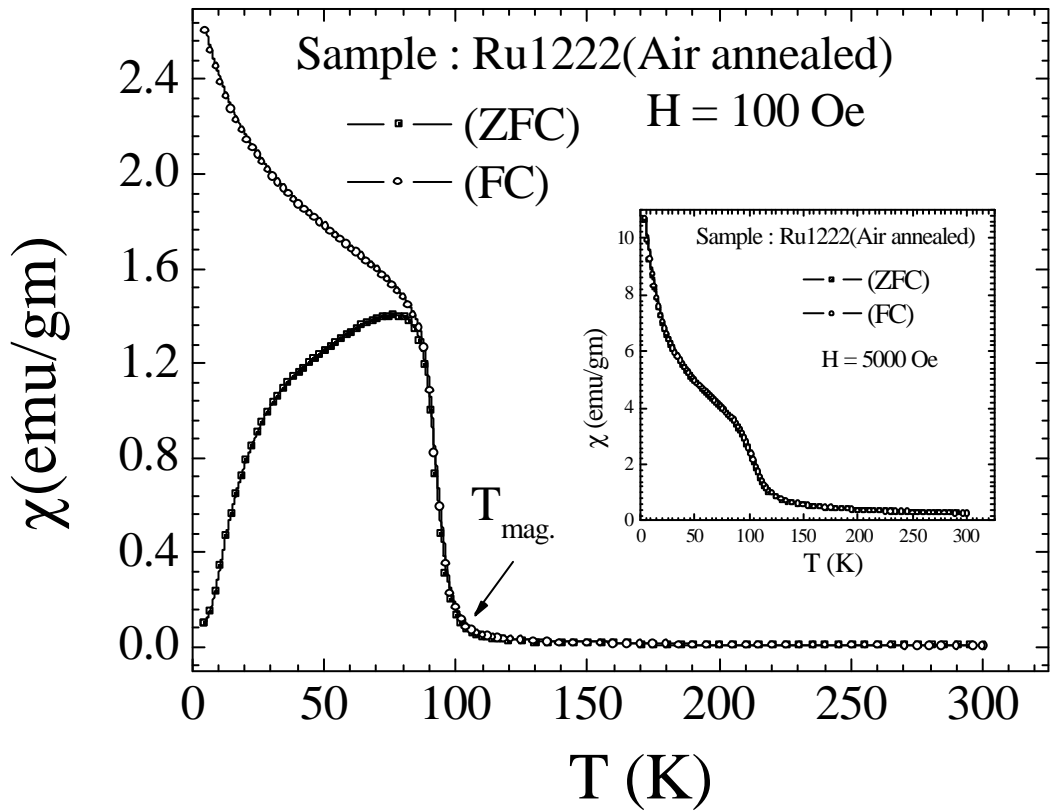

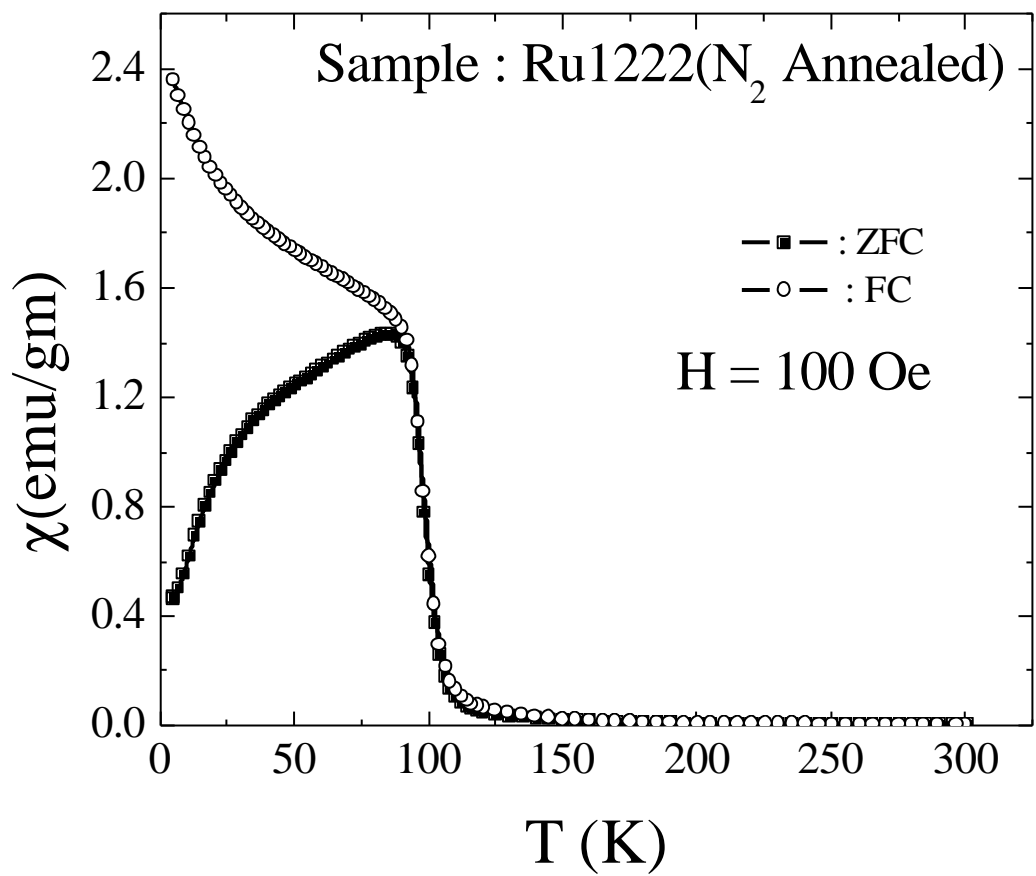

Fig.5 Awana etal.

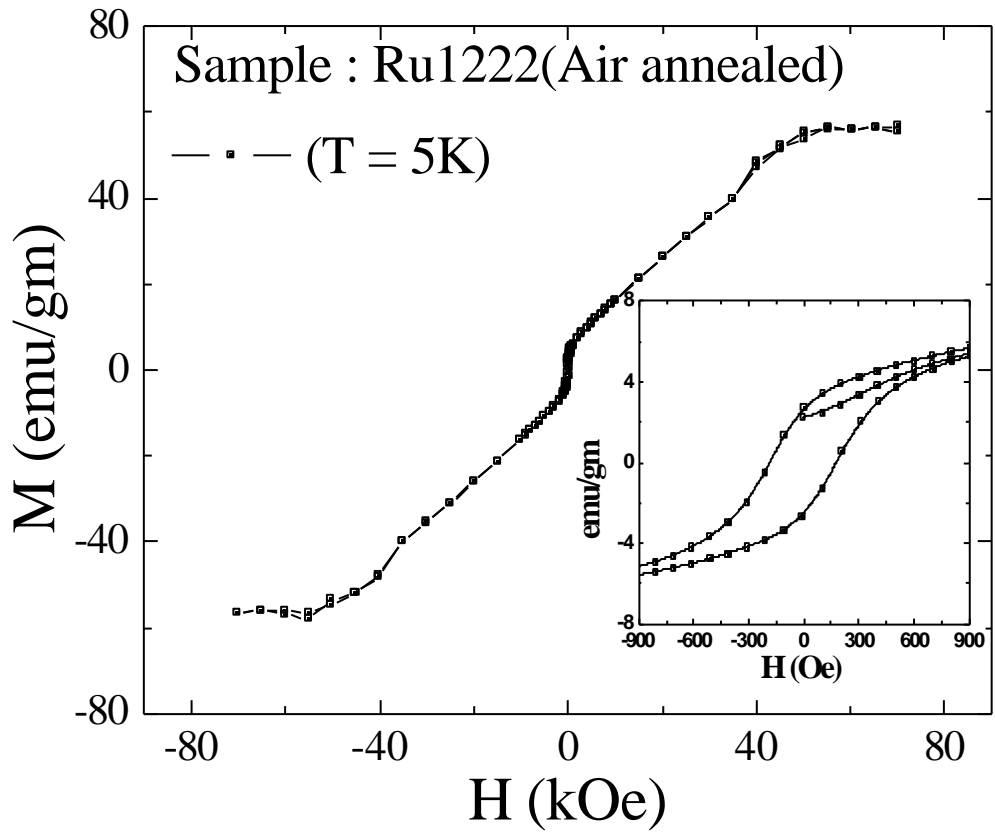

Fig.7 Awana etal.

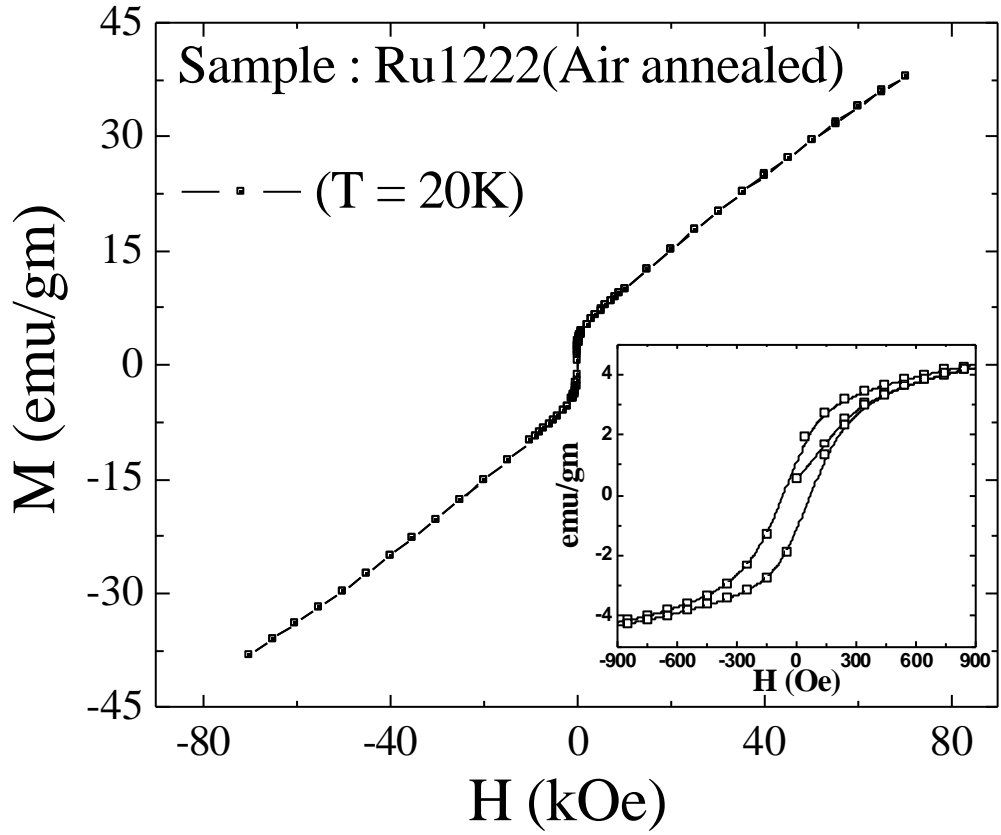

Fig.8 Awana etal.

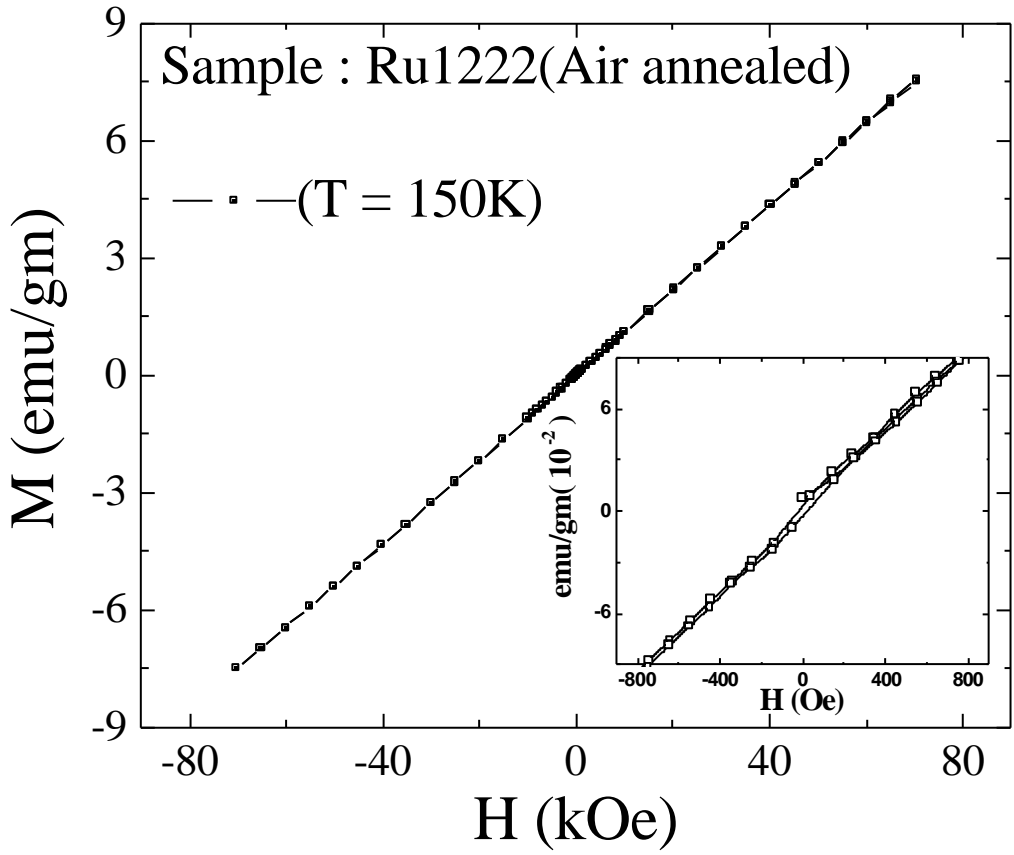

Fig.9 Awana etal.

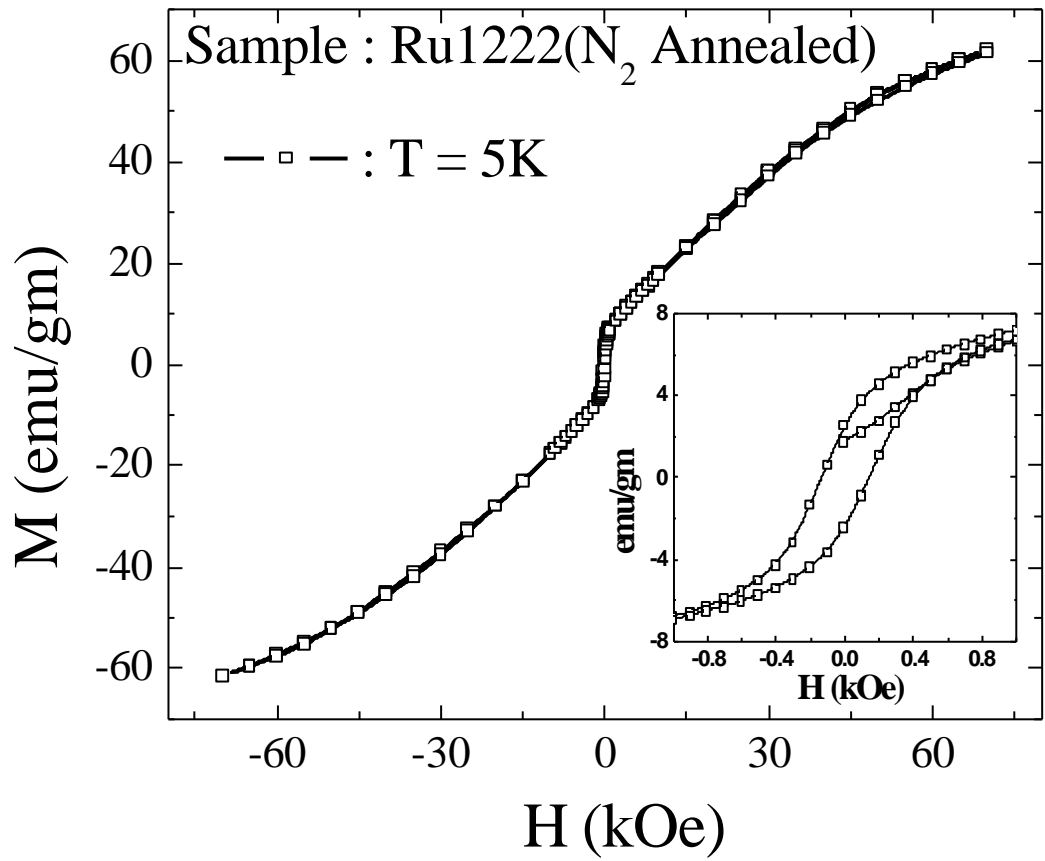

Fig.10 Awana etal.

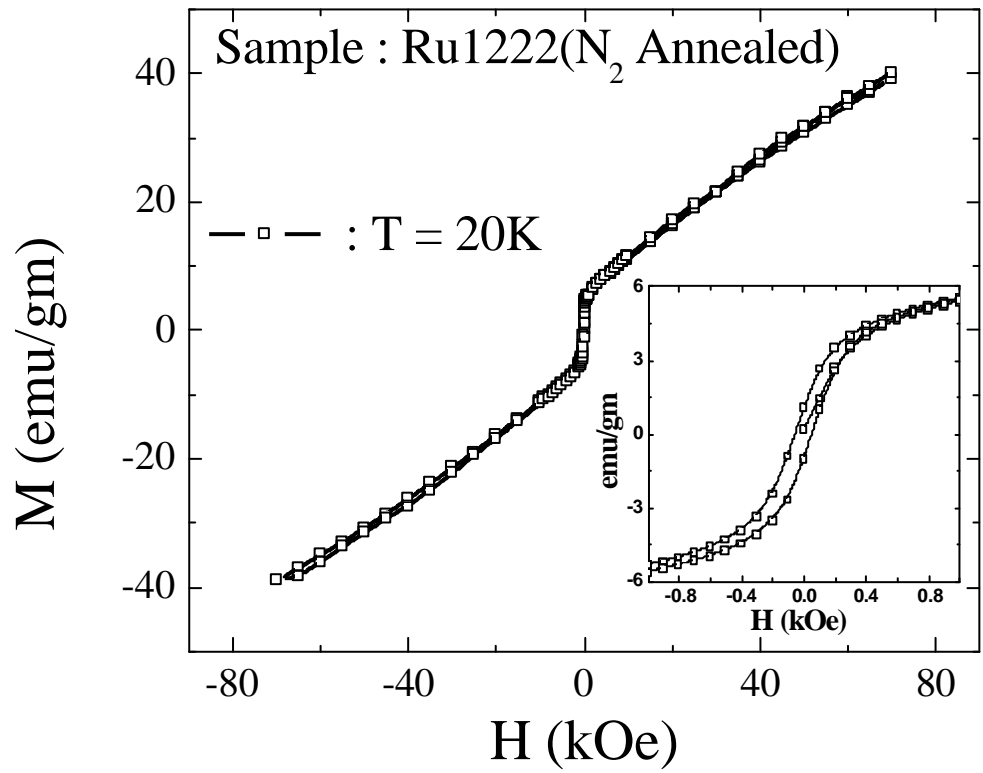